%% file: main.tex
\documentclass[10pt,twocolumn,letterpaper]{article}

\usepackage{iccv}
\usepackage{multirow}
\usepackage{colortbl} 
\usepackage{float} 
\usepackage{amssymb} 
\usepackage{bbding} 
\usepackage{gensymb}

\newcommand{\cmark}{\Checkmark} 
\newcommand{\xmark}{\XSolidBrush}

\definecolor{iccvblue}{rgb}{0.21,0.49,0.74}
\usepackage[pagebackref,breaklinks,colorlinks,allcolors=iccvblue]{hyperref}

\title{AV-Surf: Surface-Enhanced Geometry-Aware Novel-View Acoustic Synthesis}

\author{
Hadam Baek \textsuperscript{\normalfont 1} \quad
Hannie Shin \textsuperscript{\normalfont 1} \quad
Jiyoung Seo\textsuperscript{\normalfont 1}\quad
Chanwoo Kim \textsuperscript{\normalfont 1} \quad \\
Saerom Kim \textsuperscript{\normalfont 1}\quad
Hyeongbok Kim \textsuperscript{\normalfont 2} \quad 
Sangpil Kim \textsuperscript{\normalfont 1}\thanks{Corresponding author.}\vspace{0.6em}\\
\textsuperscript{\normalfont 1}Korea University \quad
\textsuperscript{\normalfont 2}Testworks \quad
\vspace{-1.em}
}

\begin{document}
\maketitle
\input{tex/abstract.tex}

\input{tex/introduction.tex}
\input{tex/relatedwork.tex}

\input{tex/method.tex}
\input{tex/experiment.tex}
\input{tex/conclusion.tex}
{
    \small
    \bibliographystyle{ieeenat_fullname}
    \bibliography{main}
}

\end{document}

%% file: tex/abstract.tex
\begin{abstract}
Accurately modeling sound propagation with complex real-world environments is essential for Novel View Acoustic Synthesis (NVAS).
While previous studies have leveraged visual perception to estimate spatial acoustics, the combined use of surface normal and structural details from 3D representations in acoustic modeling has been underexplored.
Given their direct impact on sound wave reflections and propagation, surface normals should be jointly modeled with structural details to achieve accurate spatial acoustics.
In this paper, we propose a surface-enhanced geometry-aware approach for NVAS to improve spatial acoustic modeling.
To achieve this, we exploit geometric priors, such as image, depth map, surface normals, and point clouds obtained using a 3D Gaussian Splatting (3DGS) based framework.
We introduce a dual cross-attention-based transformer integrating geometrical constraints into frequency query to understand the surroundings of the emitter.
Additionally, we design a ConvNeXt-based spectral features processing network called Spectral Refinement Network (SRN) to synthesize realistic binaural audio.
Experimental results on the RWAVS and SoundSpace datasets highlight the necessity of our approach, as it surpasses existing methods in novel view acoustic synthesis.
\footnotetext{Project Page: \href{https://avsurf.github.io/}{https://avsurf.github.io/}}
\vspace{-1em}
\end{abstract}

%% file: tex/introduction.tex
\section{Introduction}

\vspace{-0.5em}

Novel View Acoustic Synthesis (NVAS) enables a more realistic auditory experience by reconstructing accurate physically modeled spatial audio from a novel emitter position.
The advancement of NVAS provides immersive reality to various spatial audio-requiring industries, including augmented reality (AR), virtual reality (VR), gaming, and more.
However, despite advances in neural acoustic modeling, achieving  realistic spatial audio in real-world scenes remains challenging. 
Unlike controllable virtual environments, real-world acoustic synthesis requires a detailed understanding of how sound interacts with complex geometry, including indoor architectural structures and irregular obstacles.

\noindent Many existing methods~\cite{ratnarajah2024av,liang2023av, luo2022learning} use geometric priors to model acoustics, but they often rely on incomplete assumptions that do not entirely capture real-world environments' complexity.
As a result, these approaches struggle to adapt to different settings, leading to errors in simulating realistic reverberation, occlusion, and spatial sound diffusion.
To address these challenges, it is essential to integrate richer 3D geometry information into acoustic models to better capture spatial relationships in complex real-world environment.
Meanwhile, recent studies~\cite{liang2023av,bhosale2024av} have explored approaches incorporating visual cues to enhance geometric modeling for spatial acoustics.

\begin{figure}[t]
    \centering
    \includegraphics[width=\linewidth]{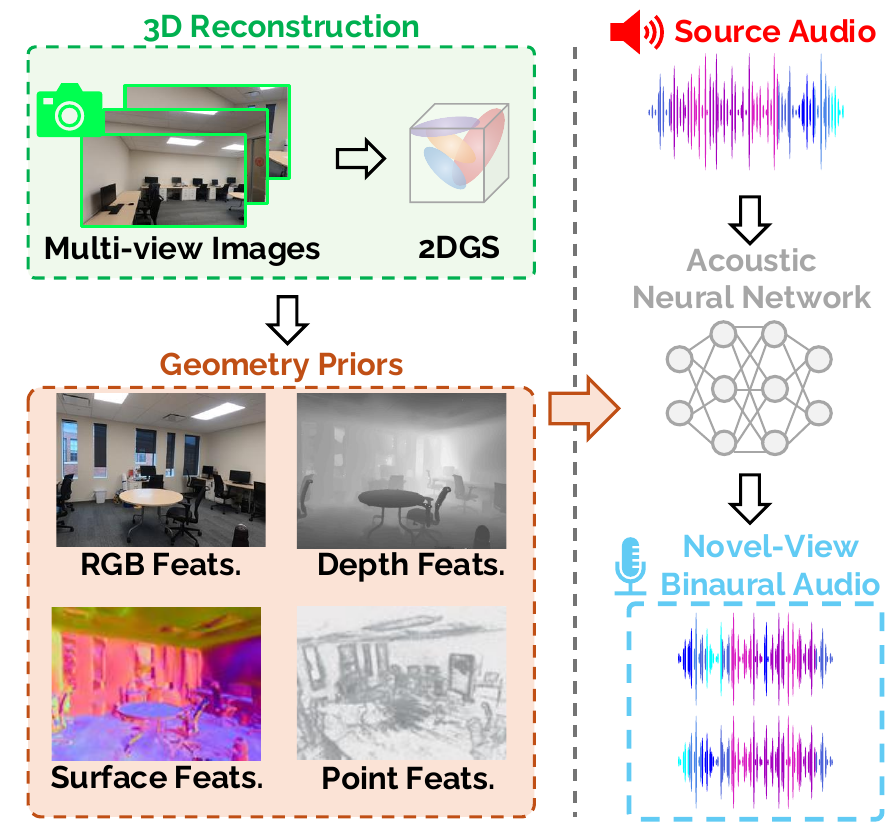}
    \vspace{-2.em}
    \caption{
    Our proposed AV-Surf learns how acoustics interact with various geometry cues in the real world and uses them to predict sound source leads to binaural audio from a novel-viewpoint.
    }
    \label{fig:intro}
    \vspace{-2em}
\end{figure}

The structural architecture of geometry predominantly influences how sound propagates in real-world environments as they determine how sound waves travel, reflect and diffract across a scene.~\cite{GROVES1955113}
However, surface properties also play a significant role in fine-grained acoustic interactions and cannot be ignored, influencing directional reflections~\cite{ATTENBOROUGH1985521}, micro-diffusions, and material-dependent absorption.
This suggests that combining structural form with surface properties can enable more realistic spatial acoustic modeling.
Accurately capturing these effects in real-world acoustics requires reliable and precise geometric information.
Recently, 3D Gaussian Splatting~\cite{kerbl20233d} (3DGS) has emerged as a powerful approach for representing complex 3D scenes while providing essential geometrical information for modeling spatial acoustics.
However, 3DGS inherently suffers from multi-view surface inconsistencies due to its use of 3D Gaussians. 
2DGS~\cite{huang20242d} resolves this issue by leveraging 2D Gaussian representations.
Accordingly, we utilize 2DGS to extract surface normal information from the 3D representation.

\noindent To effectively integrate these multimodal cues, we design a Transformer that processes and fuses visual information, surface-aware features, and 3D geometric priors, following previous works~\cite{roh2024catsplat,roh2024edge} on multimodal fusion.
Our Transformer framework learns to combine these representations by employing dual cross-attention~\cite{vaswani2017attention} mechanisms, enabling a more accurate and spatially consistent acoustic model that generalizes across diverse real-world scenes.
Unlike previous methods that rely solely on geometric priors or limited visual cues, our approach directly utilizes visual features, surface normals, and 3D geometry in a complementary manner to enhance spatial audio synthesis.

\noindent
To further enhance the quality of binaural audio synthesis with the geometry-integrated acoustics gained by our transformer, we develop a spectral refine module based on ConvNeXt~\cite{liu2022convnet,woo2023convnext}.
We apply multi-kernel architecture to enable model can learn frequency relationships across a broader range of frequencies in the spectrogram while preventing amplitude distortion.
Compared to conventional architectures~\cite{richard2021neural}, our spectral refinement network designed to capture more fine-grained time-frequency energy. 

\noindent 
In Fig.\ref{fig:intro} we provide an overview of our approach to synthesizing realistic binaural audio. 
Using the challenging real world dataset RWAVS~\cite{liang2023av}, we demonstrate the superiority of AV-Surf in generating realistic binaural audio.
Furthermore, we provide a detailed look at how our AV-Surf generates more realistic binaural audio to demonstrate the importance of surface normal for modeling spatial acoustics and spectral refine network.

\noindent Our key contributions are as follows:

\begin{itemize}

\item 
We introduce a Transformer-based feature fusion method that injects visual, geometric, and surface area cues into frequency embeddings to embed the rich features of real-world environments that influence sound propagation.

\item 
We propose a ConvNeXt-based spectral feature enhancing module that produces realistic binaural audio by refining features in various receptive field from spectrogram in the time-frequency domain. 

\item Through extensive evaluation on real world dataset and simulation dataset, RWAVS and SoundSpaces, our work outperforms existing methods in Novel View Acoustic Synthesis.

\end{itemize}

%% file: tex/relatedwork.tex
\section{Related Work}
\vspace{-0.5em}
\paragraph{Spatial Audio Binauralization.}

Audio binauralization aims to convert monaural audio signals into binaural signals, enhancing spatial perception by simulating sound directionality, distance, and spatial cues. Generating realistic binaural audio from mono sources is a challenging task, and various signal processing approaches ~\cite{antonello2017room} have been explored to improve spatial audio synthesis from mono signals.
Serr`a et al.~\cite{serra2023mono} proposed a method that predicts parametric stereo parameters from mono signals and reconstructs stereo signals using a decoder. However, the issue of temporal consistency arises as certain instruments unexpectedly shift between channels, requiring an expanded receptive field or hierarchical learning-based model to address it. 
Several Digital Signal Processing(DSP)-based methods ~\cite{zotkin2004rendering, jianjun2015natural} synthesize binaural audio by integrating Room Impulse Response (RIR) ~\cite{antonello2017room, kelley2024rir} and Head-Related Transfer Functions (HRTF) ~\cite{cheng2001introduction}, both of which are costly to measure due to the need for specialized equipment and individualized calibration. 
Recent studies have proposed neural network-based learning methods to generate more realistic binaural audio. Richard et al.~\cite{richard2021neural} introduced a ConvNet-based binauralization model designed with convolutional layers, capturing finer audio characteristics, while Leng et al.~\cite{leng2022binauralgrad} proposes a two-stage framework utilizing diffusion models to generate binaural audio and employs a diffusion denoising decoder to learn sound patterns. 
In comparison, our approach proposed a ConvNeXt~\cite{woo2023convnext}-based acoustic binauralizer, which captures more comprehensive spectral characteristics of the audio.

\vspace{-2.2em}
\paragraph{Surface-Aware Geometric Priors with 3D Gaussian Splatting.}
Recently, 3D Gaussian Splatting (3DGS) has emerged and has been rapidly adopted across multiple domains. Several studies have attempted to accurately extract rich geometric features such as surface normals and depth from explicit 3DGS scene representation, a task that remains highly challenging. 
Previous works ~\cite{yang2022neumesh, darmon2022improving} optimize signed distance functions (SDF) for implicit neural mesh-based fields, then Marching Cube algorithm ~\cite{lorensen1998marching} is used to reconstruct triangle mesh. 
SuGaR~\cite{guedon2024sugar} proposed regularization term and utilizes Poisson reconstruction~\cite{kazhdan2006poisson} to obtain mesh-based representation from 3DGS, which is itself much faster than NeRF~\cite{mildenhall2021nerf}. Gaussian Opacity Fields ~\cite{yu2024gaussian} and RaDe-GS~\cite{zhang2024rade} encourage the Gaussians to be well spread which aligned more closely to the surface by leveraging the ray-tracing Gaussian intersection method. Unlike previous works, 2DGS~\cite{huang20242d} directly employs 2D Gaussians, enhancing the resulting surface geometry without additional refinement strategy. 
To the best of our knowledge, no existing method utilizes surface-aware geometric prior for real-world acoustic modeling. 

\begin{figure*}[t]
    \centering
    \includegraphics[width=\linewidth]{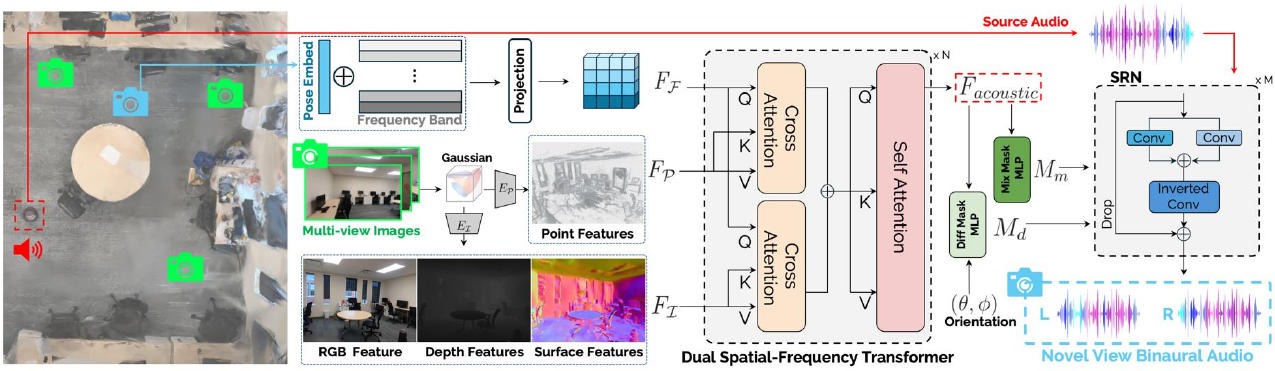}
    \vspace{-2.em}
    \caption{
    \textbf{AV-Surf overview.} 
    AV-Surf first trains 2DGS with given multi-view images to obtain various types of real-world environment information. Then we utilize two types of encoders, \(\mathit{E}_{\mathcal{P}}\) and \(\mathit{E}_{\mathcal{I}}\), to extract geometry and spatial features, \(\mathit{F}_\mathcal{P}\) and \(\mathit{F}_\mathcal{I}\). We inject spatial cues to position added frequency embeddings \(\mathit{F}_\mathcal{F}\) with iterative transformer layers to learn real world acoustics. After \(\mathit{F}_{acoustic}\) is obtained, AV-Surf follows the method from the previous study~\cite{liang2023av} to get mixture \(\mathit{M}_m\) and difference \(\mathit{M}_d\) acoustic masks. 
    Finally, our ConvNeXt-based Spectral Refinement Network (SRN)  takes the sound source and both masks as inputs, refining them to generate realistic novel-view binaural audio.
    }
    \label{fig:overview}
    \vspace{-1.8em}
\end{figure*}

\vspace{-1.2em}
\paragraph{Visually Informed Neural Audio Synthesis.}

Prior works~\cite {liang2023av,luo2022learning, su2022inras, ratnarajah2024av, ahn2023novel, ratnarajah2024listen2scene} have explored various learning-based strategies and visual cues to improve spatial audio generation.
Ahn et al.~\cite{ahn2023novel} propose a room impulse response (RIR) deconvolution method, however, their performance relies on the availability and quality of RIRs and utilizes insufficient geometric priors of 3D representation. AV-NeRF~\cite{liang2023av} utilizes Neural Radiance Fields (NeRF)~\cite{mildenhall2021nerf,tancik2023nerfstudio}, which employs MLPs to learn an implicit representation of 3D scene representation and generate spatial sound, integrating the source position and encoded visual features into the pipeline using AV-mapper. Neural Acoustic Field (NAF)~\cite{luo2022learning} introduced an implicit neural representation to learn the spatial relationship between the speaker and the listener, while INRAS~\cite{su2022inras} proposed an approach that refines interactive acoustic radiance transfer by using scene geometry as input. Recently, AV-RIR~\cite{ratnarajah2024av} introduced scene-aware binaural sound simulation by integrating geometric topology to improve accuracy in sound propagation modeling. 
In contrast to the implicit neural 3D representation learning, 3DGS ~\cite{kerbl20233d} utilizes discrete Gaussians to represent scenes with high spatial expressiveness, enabling high-quality real-time rendering without the need for neural network forward propagation. 
Listen2Scene~\cite{ratnarajah2024listen2scene} introduce sound propagation architecture that generates BIR (Binaural Impulse Response) from 3D mesh-based representation learning. 
Recently, AV-GS ~\cite{bhosale2024av} focuses on learning audio-visual representations that are conditioned on explicit 3D Gaussian representation, enabling controllable generation. However, the direction and distance awareness approach has limitations in sufficiently capturing the effects of sound attenuation influenced by geometry of the 3D environment. 
Unlike previous studies that rely on implicit geometric representations~\cite{liang2023av, luo2022learning} or predefined material properties~\cite{bhosale2024av}, our approach aims to bridge the gap between visual features and spatial audio generation by utilizing enriched geometric priors that fused with frequency embeddings.

%% file: tex/method.tex
\section{Method}
\vspace{-0.6em}
In this section, we introduce AV-Surf, a novel framework for binaural audio generation designed to achieve two key objectives:
(i) integrating of surface and 3D geometry priors into position-aware frequency embedding using a transformer with dual cross-attention
(ii) developing a spectral feature enhancing module by leveraging a ConvNeXt-based structure.
We begin by presenting the preliminary concepts that form the foundation of our approach, followed by an overview of the entire pipeline.
Finally, we provide a detailed explanation of the overall methodology.

\subsection{Preliminary}
\label{sec:3-1}
\vspace{-0.4em}
\paragraph{2D Gaussian Splatting.} 
2D Gaussian Splatting~\cite{huang20242d} addresses the multi-view surface inconsistency inherent in 3D Gaussians due to their 3D representation.
It utilizes 2D-oriented planar Gaussian disks to provide view-consistent geometry.
2DGS represents 3D space with set of 2D Gaussians, which is parameterized by :
\vspace{-2em}

\begin{equation}
    P(u, v) = \mathbf{p}_k + s_u \mathbf{t}_u u + s_v \mathbf{t}_v v = \mathbf{H}(u,v,1,1)^\mathbf{T}
    \vspace{-0.5em}
\end{equation}

\noindent where  \(\mathbf{p}_k\) denotes center point, \(\mathbf{t}_u\) and \(\mathbf{t}_v\) denotes tangential vectors, \(s_u\) and \(s_v\) denotes scaling vector of each single 2D Gaussians. 

\noindent The rasterization process of 2DGS similarly follows the 3DGS. In the initial stage of rendering, a screen space bounding box is computed for each Gaussian primitive. Following this, 2D Gaussians are arranged in a sorted order according to the depth of their center point. Finally, Volumetric alpha blending is applied to integrate alpha-weighted appearance from front to back.

\noindent The iterative process ends when the accumulated opacity is saturated. More details on 2DGS can be found in ~\cite{huang20242d}.

\vspace{-1.6em}
\begin{small}
\begin{equation}
    \mathbf{c}(\mathbf{x}) = \sum_{i=1} \mathbf{c}_i \alpha_i \hat{\mathcal{G}}_i(\mathbf{u}(\mathbf{x}))\prod^{i-1}_{j=1}(1-\alpha_j\hat{\mathcal{G}}_j(\mathbf{u}(\mathbf{x})))
    \vspace{-0.5em}
\end{equation}
\end{small}

\subsection{Overview}
\label{sec:3-2}
\vspace{-0.5em}
Our proposed framework, AV-Surf, illustrated in Fig.\ref{fig:overview} follows a structured approach to generate high-quality binaural audio by leveraging spatial and geometric cues. We first train the aforementioned 2DGS with initialized SFM~\cite{snavely2006photo} point clouds from given multi-view images to reconstruct a 3D scene representation. From this reconstructed 3D structure, we directly extract 3D points from each Gaussian and sample points \(\mathcal{P} \in \mathbb{R}^{N \times 3}\) based on ~\cite{qi2017pointnet++}. Moreover, we render views from Gaussians, including 2D image \(\mathcal{I}_v \in \mathbb{R}^{W\times H \times 3}\), depth \(\mathcal{I}_d \in \mathbb{R}^{W\times H \times 1}\), and surface normal \(\mathcal{I}_n \in \mathbb{R}^{W\times H \times 3}\).
We then concatenate 2D spatial cues \{\(\mathcal{I}_v,\mathcal{I}_d,\mathcal{I}_n\)\} along the channel dimension and process them through ResNet\cite{he2016deep}-based encoder to obtain 2D feature \(\mathit{F}_{\mathcal{I}} \in \mathbb{R}^{W' \times H' \times D_i} \), while \(\mathcal{P}\) are processed using PointNet\cite{qi2017pointnet} to extract geometric feature \(\mathit{F}_{\mathcal{P}} \in \mathbb{R}^{N' \times \ D_p}\). 
These extracted features are aligned with position-combined frequency embeddings \(\mathit{F}_{\mathcal{F}} \in \mathbb{R}^{N_{\mathcal{F}} \times D_f}\) (\(N_{\mathcal{F}}\) denotes the number of frequency bins) via our transformer module to integrate geometrical representation.
This process allows us to obtain a geometry-aware frequency feature,  \(\mathit{F}_{acoustic}\).
To utilize  \(\mathit{F}_{acoustic}\), we adopt the method described in ~\cite{liang2023av}  to compute the mixture and difference magnitudes of the spectrogram are obtained by performing the Short-Time Fourier Transform (STFT) on the sound source waveform.
Then, we generate novel-view binaural audio using our newly designed audio processing module, which leverages ConvNeXt~\cite{woo2023convnext} to process spectral features, enhancing the accuracy of the synthesized audio.
Through this process, our approach enhances binaural audio synthesis by leveraging geometric constraints and spectral refinement.

\begin{figure}[t]
    \centering
    \includegraphics[width=\linewidth]{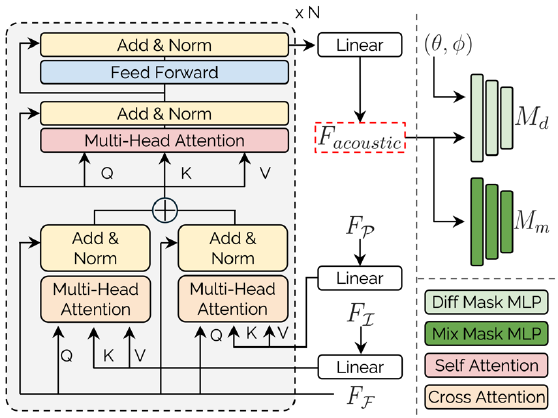}
    \vspace{-2.em}
    \caption{
    We project the spatial-geometric features to frequency dimension and carefully integrated into position aware frequency embeddings by our Transformer architecture. Spatial-geometry aligned \(\mathcal{F}_{acoustic}\) feature is then delivered to MLP for generating mixture mask and difference mask.
    }
    \label{fig:method_tf}
    \vspace{-1.8em}
\end{figure}

\vspace{-0.6em}
\subsection{Dual Spatial-Frequency Fusion Transformer}
\label{sec:3-3}
\vspace{-0.5em}
Transformer-based architectures have demonstrated strong capabilities in multi-modal feature integration~\cite{roh2024catsplat}, making them particularly well-suited for cross-domain fusion. 
Inspired by these advances, we propose a geometry-aware acoustic feature representation that effectively encodes spatial cues for binaural audio synthesis. 
Our method employs a Transformer-based fusion module that sequentially integrates spatial geometric features to position-aware frequency embeddings, ensuring a spatially aligned acoustic representation.
To achieve spatial-aware acoustic feature learning, we employ a dual cross-attention mechanism followed by self-attention and feed-forward refinement.
This mechanism integrates \(F_\mathcal{I}\) and \(F_\mathcal{P}\) into \(F_\mathcal{F}\), ensuring a more effective feature interaction.
Instead of a single directional attention operation, both \(F_\mathcal{I}\) and \(F_\mathcal{P}\) contribute to refining \(F_\mathcal{F}\) by interacting in two parallel attention streams.
For each fusion step, the query matrix \(\mathbf{Q}\) is derived from the frequency feature \(F_\mathcal{F}\), while the key and value matrices \(\mathbf{K}\), \(\mathbf{V}\) originate from the respective integrating feature (\(F_\mathcal{I}\) or \(F_\mathcal{P}\)). This is formulated as:
\vspace{-0.8em}
\begin{equation}
    \mathbf{Q}=W_QF_\mathcal{F}, \mathbf{K}=W_KF', \mathbf{V}=W_VF'
    \vspace{-0.8em}
\end{equation}
where \(W_Q\), \(W_K\), \(W_V\) are learnable projection matrices, and \(F'\) denotes the interacting feature (\(F_\mathcal{I}\) or \(F_\mathcal{P}\)).
We then compute cross attention using projected \(\mathbf{Q}\), \(\mathbf{K}\), and \(\mathbf{V}\) with the following formula:
\vspace{-0.8em}
\begin{equation}
    \mathbf{Attn}_{F'}=\text{softmax} \left(\frac{\mathbf{Q} \mathbf{K}^{T}}{\sqrt{D_f}} \right)
    \mathbf{V}
    \vspace{-0.8em}
\end{equation}
where \(D_f\) denotes the size of frequency feature dimension. The outputs of both attention modules are combined and integrated into the frequency feature through residual connections and layer normalization:
\vspace{-0.8em}
\begin{equation}
    F_{sum} = \text{LayerNorm}(F_\mathcal{F}+\mathbf{Attn}_{F_\mathcal{I}}+\mathbf{Attn}_{F_\mathcal{P}})
    \vspace{-0.8em}
\end{equation}
This ensures that \(F_\mathcal{I}\) and \(F_\mathcal{P}\) contribute complementary information to \(F_\mathcal{F}\), leading to a more expressive representation.
After fusing \(F_\mathcal{I}\) and \(F_\mathcal{P}\) into \(F_\mathcal{F}\), self-attention further refines the representation by capturing relationships within the fused features:
\vspace{-0.6em}
\begin{equation}
    \mathbf{Attn}_{self} = \text{softmax}\left(\frac{\mathbf{Q}_{F_{sum}}
    \mathbf{K}_{F_{sum}}^{T}}{\sqrt{D_f}}\right)\mathbf{V}_{F_{sum}}
    \vspace{-0.4em}
\end{equation}

The final feed-forward network (FFN) consisting of two linear layers with an activation layer in between refines the feature representation, followed by a skip connection and normalization to obtain \(F_{acoustic}\). This process is formulated as:
\vspace{-0.8em}
\begin{equation}
    F_{acoustic} = \mathcal{H}(\text{LayerNorm}(F_{SA}+\text{FFN}(F_{SA})))
    \vspace{-0.8em}
\end{equation}
where \(F_{SA}\) denotes the output from the self-attention module and  \(\mathcal{H}\) is a linear layer. The FFN consists of two linear transformations with an intermediate ReLU activation, expressed as:
\vspace{-0.8em}
\begin{equation}
    \text{FFN}(F)=\mathcal{H}_2(\sigma(\mathcal{H}_1(F)))
    \vspace{-1em}
\end{equation}
where \(\mathcal{H}_1\) and \(\mathcal{H}_2\) are linear layers, and \(\sigma\) represents the ReLU activation function.
The resulting \(\mathit{F}_{acoustic}\), which is a frequency feature enriched with spatial-geometric information through the transformer, is used to generate a mixture mask independently via an MLP or a difference mask with the inclusion of additional orientation embedding.
Full procedure is illustrated in Fig.\ref{fig:method_tf}.

\vspace{-0.2em}
\subsection{Spectral Refinement Network~(SRN)}
\label{sec:3-4}
\begin{figure}[t]
    \centering
    \includegraphics[width=\linewidth]{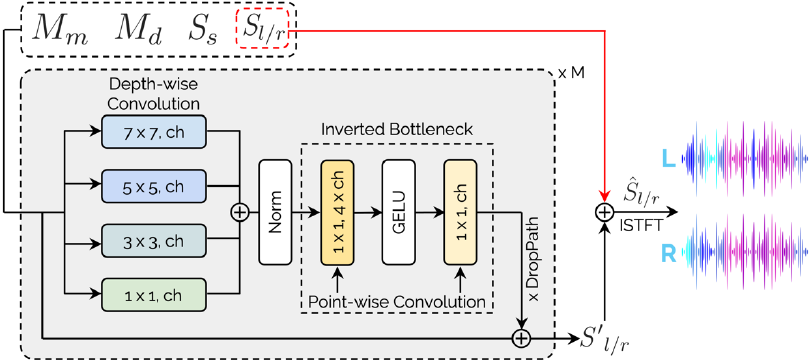}
    \vspace{-2.em}
    \caption{
    \textbf{SRN} estimates waveform based on mixture mask \(M_m\), difference mask \(M_d\), STFT spectrogram \(S_s\), estimated left and right channel spectrogram \(S_{l/r}\) by using sequential depth-wise convolution, normalization, point-wise convolution, GELU activation and DropPath.
    }
    \label{fig:method_decoder}
    \vspace{-1.9em}
\end{figure}

\vspace{-0.5em}
To generate binaural audio, we build on the existing method~\cite{liang2023av} that utilizes the sound source, the mixture mask, and the difference mask obtained from the previous step.
While approaches based on traditional convolutional audio processing are simple and effective, they lack the ability to reconstruct fine-grained audio details that directly reflect spectral representations embedded with spatial information.
To address this limitation, we propose a novel audio processing network tailored for this task.
Motivated by this insight, we design our \textbf{Spectral Refinement Network (SRN)} based on ConvNeXt, a modernized convolutional network with Transformer-inspired architectural modification that has demonstrated superior performance across various tasks.
To further enhance the audio reconstruction with fine-grained frequency-time details, we incorporate multiple receptive fields of varying sizes, ensuring a more comprehensive representation of the spatial and spectral characteristics of the audio.
Our audio processing module takes as input \(F_\mathcal{S} \in \mathbb{R}^{4 \times N_{\mathcal{F}}\times N_{\mathcal{T}}}\), a concatenation of the mixture mask \(M_m\), the difference mask \(M_d\), the source spectrogram \(S_s\), and the estimated left/right spectrogram \(S_{l/r}\).
This input undergoes a multi-kernel depthwise convolution to extract frequency-time patterns across different receptive fields:
\vspace{-0.7em}
\begin{equation}
    F_{\mathcal{S}}' = \sum_{k \in {1,3,5,7}} \mathcal{DC}(F_{\mathcal{S}}, k)
    \vspace{-0.8em}
\end{equation}
where \(\mathcal{DC}\) denotes depthwise convolution using each kernel size \(k\) extracts spectral features at different scales, and the outputs are aggregated via summation.
After convolution, the feature map undergoes layer normalizaton for improved numerical stability and convergence.
The normalized feature is then processed through an inverted bottleneck block, consisting of a pointwise channel expansion convolution, GELU~\cite{hendrycks2016gaussian} activation function, and a pointwise channel reduction convolution, with DropPath mehod~\cite{huang2016deep} to obtain refined spectral feature:
\vspace{-0.8em}
\begin{equation}
S_{l/r}' = \gamma \times\mathcal{PC}_2(\sigma_g(\mathcal{PC}_1(F_{\mathcal{S}}'))) + F_{\mathcal{S}}
\vspace{-0.6em}
\end{equation}
where \(\mathcal{PC}\) denotes pointwise convolution, \(\sigma_g\) denotes GELU activation function, and \(\gamma\) denotes DropPath ratio. This entire process is performed iteratively to obtain spectral feature \(S'_{l/r}\).
Final waveform is processed by applying Inverse Short-Time Fourier Transform (ISTFT) with following Griffin-Lim algorithm~\cite{griffin1984signal} to \(\hat{S}_{l/r}\) computed with summation of \(S'_{l/r}\) and \(S_{l/r}\). 
The procedure of our SRN is illustrated in Fig.\ref{fig:method_decoder}.

\vspace{-0.3em}
\subsection{Learning objective}
\label{sec:3-5}
\vspace{-0.5em}
Our optimization process consists of two distinct stages.
We first follow the original paper~\cite{huang20242d} to optimize a 3D scene from an initialized sparse point cloud using a set of posed images. The loss function used in this stage is as follows:
\vspace{-0.8em}
\begin{equation}
    \mathcal{L} = \mathcal{L}_c + \alpha\mathcal{L}_d + \beta\mathcal{L}_n
    \vspace{-0.8em}
\end{equation}

\noindent where \(\mathcal{L}_c\) is a RGB reconstruction loss, which combines \(\mathcal{L}_1\) with the D-SSIM term from ~\cite{kerbl20233d}, while depth distortion loss \(\mathcal{L}_d\) and normal consistency loss \(\mathcal{L}_n\) act as regularization terms defined in 2DGS.
After completing 2DGS optimization, we optimize our model by minimizing the following loss function:
\vspace{-0.6em}
\begin{small}
\begin{equation}
    \mathcal{L} = MSE(\hat{S}_M,S_M) + MSE(\hat{S}_L,S_L) + MSE(\hat{S}_R,S_R)
    \vspace{-1.2em}
\end{equation}
\end{small}
\vspace{0.8em}
\noindent where \(S\) represents the magnitude of the spectrogram obtained by performing the Short Time Fourier Transform (STFT) on sound source signals and \(\hat{S}\) represents the synthesized spectrogram output generated by our model.
\(S_M\) denotes the mono spectrogram magnitude, computed by average of the left \(S_L\) and right \(S_R\) channel. Entire loss computed using MSE loss.

%% file: tex/experiment.tex
\input{tab/rwavs_table}

\section{Experiments}
\vspace{-0.6em}
\subsection{Datasets}
\vspace{-0.3em}
\paragraph{RWAVS.}
The RWAVS~\cite{liang2023av} dataset is a real-world dataset that offers realistic binaural audios and corresponding images, camera poses. The dataset consists of 13 scenes covering three indoor environments(office, house, apartment) and one outdoor environment. For each scene, It provides realistic binaural audio and frames captured at 1 frame per second. The camera poses are extracted using COLMAP~\cite{schonberger2016structure}, and the dataset provides both the original audio, which ranges from 10 to 25 minutes in length, and the corresponding binaural audio. To maintain consistency in data structure and evaluation, the train-validation split follows the same partitioning strategy as AV-NeRF.
\vspace{-1.6em}

\paragraph{SoundSpaces.}
SoundSpaces~\cite{chen2020soundspaces} is a synthetic dataset that provides Binaural Room Impulse Responses (RIRs) considering the effects of room geometry. It includes the coordinates of both the receiver and the sound source, along with binaural RIRs corresponding to four different head orientations of the receiver (0\degree, 90\degree, 180\degree, and 270\degree). This benchmark has been widely used in IR-related studies such as ~\cite{su2022inras,luo2022learning}, enabling performance evaluation under the same conditions as previous IR studies. By analyzing the reverberation characteristics, we select six indoor scenes from the Replica dataset, which consists of two small scenes, two medium-sized scenes, and two large scenes. The dataset is provided through Habitat Sim~\cite{savva2019habitat}, an audio simulator, which enables the extraction of visual information and audio using the simulator. We follow the same training/test split as previous works~\cite{luo2022learning} by using 90\% for training and 10\% data for testing.

\input{tab/soundspace_tab.tex}

\subsection{Evalutaion Metrics}
\vspace{-0.4em}
To evaluate different methods, we use distinct metrics for the RWAVS and SoundSpaces datasets. 
For the RWAVS dataset, we assess performance using magnitude distance (MAG)~\cite{xu2021visually}, which evaluates audio quality in the time-frequency domain, and envelope distance (ENV)~\cite{morgado2018self}, which measures audio quality in the time domain. 
Meanwhile, for the SoundSpaces dataset, following previous studies~\cite{su2022inras,luo2022learning}, we evaluate the predicted impulse responses using three key acoustic metrics: Reverberation Time (T60), Clarity (C50), and Early Decay Time (EDT).
T60 quantifies the overall reverberation by measuring how long it takes for an impulse response to decay by 60 dB.
C50 represents the energy ratio within the first 50 ms of the RIR to the remaining portion, directly influencing speech intelligibility and acoustic clarity.
EDT captures the characteristics of early reflections.

\vspace{-0.5em}
\subsection{Implementation Details}
\vspace{-0.5em}
Our model is implemented with PyTorch~\cite{paszke2019pytorch} using a single Nividia RTX 4090 GPU. 
We followed the training setup of 2DGS and provide a detailed desciption of our remaining model implementation.
We train our model with Adam~\cite{kingma2014adam} optimizer with learning rate 1e-4.
For the encoder that extracts features of priors from 2DGS, we used ResNet-18 with its last three layer removed.
Additionally, we utilized PointNet with its global pooling layer removed. 
We employed four transformer blocks each with 256 feature dimensions and 8 heads for both cross attention and self attention. 
Following previous research~\cite{liang2023av}, we employed MLP~\cite{popescu2009multilayer} consisting of 4 linear layers following ReLU activation~\cite{banerjee2019empirical} except the last layer to obtain mixture and difference masks.
For SRN, each depthwise convolution has 16 channels, and we used 9 SRN blocks. We trained our model for 50 epochs.

\begin{figure*}

    \centering
    \includegraphics[width=\linewidth]{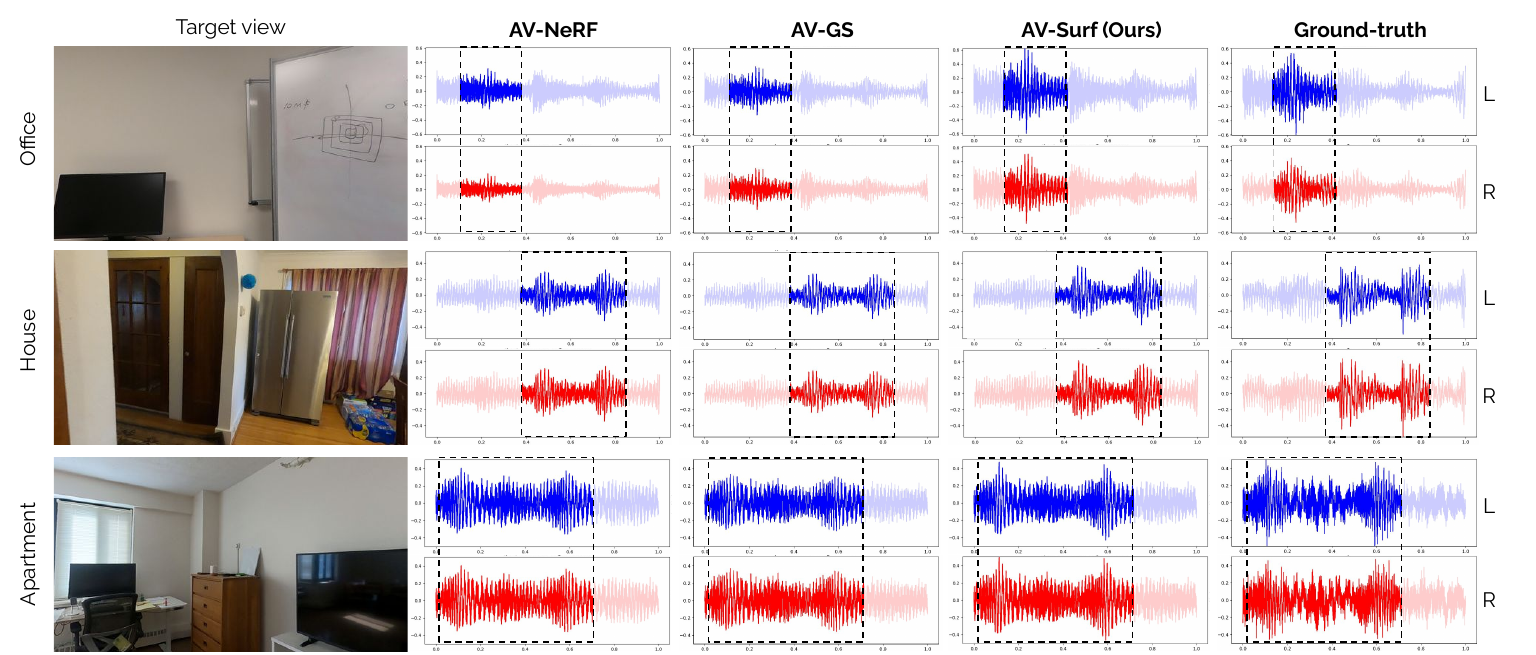
    }
    \vspace{-2.em}
    \caption{
    Qualitative results on RWAVS dataset. 
    Comparison of AV-Surf, AV-NeRF, and AV-GS against the ground-truth.
    }
    \label{fig:quality}
    \vspace{-1.6em}
    
\end{figure*}

\begin{figure}[t]
    \centering
    \includegraphics[width=\linewidth]{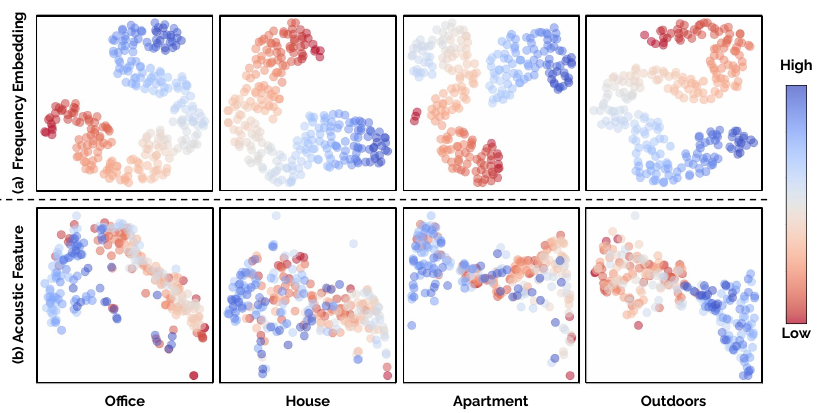}
    \vspace{-2.em}
    \caption{
    t-SNE Visualization of two types of feature, (a)frequency embeddings \(F_\mathcal{F}\), (b)acoustic feature \(F_{acoustic}\). The bar graph represents the color of the frequency bins.
    }
    \label{fig:t_sne}
    \vspace{-1.8em}
\end{figure}

\vspace{-0.4em}
\subsection{Quantitative Results}
\vspace{-0.5em}
\paragraph{Novel view binaural audio synthesis on RWAVS dataset.}
We compare our AV-Surf with the baselines tested on the real-world dataset RWAVS, as shown in Table.\ref{tab:rwavs}. 
Mono-Mono replicates the source audio, producing fake binaural audio without any process.
Mono-Energy scales the amplitude of input audio to align with the target and duplicates to create binaural audio. 
Stereo-Energy scales the stereo input audio to match with the target.
We compare INRAS~\cite{su2022inras}, NAF~\cite{luo2022learning}, ViGAS~\cite{chen2023novel}, AV-NeRF~\cite{liang2023av}, AV-GS~\cite{bhosale2024av} for the neural network-based methods.
Compared to AV-NeRF, our model achieves substantial improvements in MAG and ENV, with a 9.7\% improvement in MAG and 3.4\% improvement in ENV, demonstrating its effectiveness in utilizing surface-integrated geometry.
Furthermore, our model achieves competitive performance in ENV compared to the state-of-the-art AV-GS model and significantly outperforms it in MAG, achieving a 4.2\% improvement.

\vspace{-1.6em}
\paragraph{Room Impulse Response on SoundSpaces dataset.}
We compare our AV-Surf with the following baselines on the synthetic dataset SoundSpaces, as shown in Table.\ref{tab:soundspaces}.
Linear and nearest neighbor interpolation methods are applied to two widely used traditional audio coding techniques, Advanced Audio Coding(AAC)~\cite{ISO13818-7} and Xiph Opus~\cite{xiph_opus}.
We compare AV-Surf with NAF~\cite{luo2022learning}, INRAS~\cite{su2022inras}, AV-NeRF~\cite{liang2023av}, and AV-GS~\cite{bhosale2024av} for neural network-based methods.
Our model demonstrates strong performance in EDT while achieving significant improvements of 8.52\% T60 and 1.89\% in C50, further highlighting its effectiveness on the SoundSpaces dataset.

\vspace{-0.4em}
\subsection{Qualitative Results}
\vspace{-0.6em}
\paragraph{Audio render enhancement.}
We visualize the binauralized audio generated through three different methods, including our AV-Surf, with ground truth in Fig.~\ref{fig:quality}.
As we can see in the figure, across all scenes, through our SRN, the refined generated AV-surf waveform captures detailed amplitude variations compared to previous methods.
AV-NeRF primarily relies on visual cues from the listener's perspective, while AV-GS leverages learned Gaussians to preserve the overall geometric structure.
Based on these two advantages, we add surface normal information to the listner's perspective visual cues and leverage geometric features extracted from 2DGS further improves the performance of generating binaural audio.
Compared to the previously mentioned methods, AV-NeRF and AV-GS, our model outperforms in generating more accurate amplitude for binauralized audio, as seen in the results of the office scene in the figure.
Our model also produces better results in other scenes, such as house and apartment scenes.
All of the results demonstrate the effectiveness of our method.

\vspace{-1.2em}

\begin{figure*}[t]
    \centering
    \includegraphics[width=\linewidth]{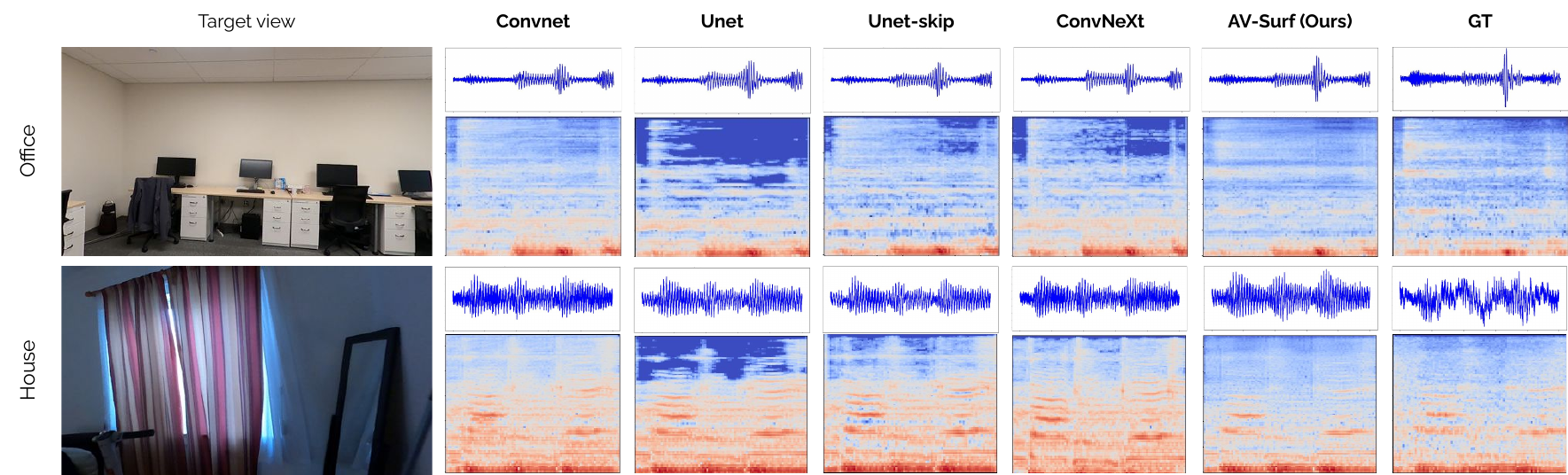}
    \vspace{-2.em}
    \caption{
       Comparison of different types of networks for refining spectral features.
    }
    \label{fig:srn}
    \vspace{-1em}
\end{figure*}

\paragraph{Learning geometric features.}
To validate our model's ability fusing spatial geometry feature into frequency space, we visualize the t-SNE distributions of frequency embeddings before transformer interaction and acoustic features after interaction.
We provide our t-SNE results in Fig.~\ref{fig:t_sne}.
Here, we observe that the initial embedding of linear frequency bands in Fig.~\ref{fig:t_sne}(a) are distinguishable. Whereas geometry-integrated acoustic features in Fig.~\ref{fig:t_sne}(b) are transformed to have overlap distributions.
This result implies that the overlap between low and mid-range frequencies indicates that these two frequency bands are primarily learning similar sound variations, while the embeddings of high frequencies, which have less overlap, demonstrate that they are learning different sound variations.
In our work, these acoustic features are utilized to obtain a mask for generating audio that better reflects a broader range of spatial acoustic characteristics.
\vspace{-0.4em}

\subsection{Ablation Study}
\vspace{-0.5em}
\paragraph{Geometric priors effect.}
In acoustic modeling, distinct geometric representations influence the results in different ways.
Surface normals and global geometric structures are fundamental to real-world sound propagation, influencing key acoustic phenomena.
To validate the impact of each factor, we design an experiment to analyze their contributions to acoustic modeling.
We present the result of this analysis in Table.\ref{tab:ablation_1_a}.
Empirically, the results demonstrate that surface normal and geometry information independently contribute to improved performance, and combining these features leads to even better results.
Since how geometry information is learned significantly affects model performance, we further investigate how to integrate these priors effectively.
As demonstrated by ~\cite{liang2023av}, using an MLP-based approach for feature alignment has proven effective.
Based on this insight, we conducted the following experiments: (1) an MLP-based fusion method, (2) a feature concatenation approach followed by cross-attention, and (3) a dual cross-attention mechanism.
As shown in Table. \ref{tab:ablation_1_b}, our approach demonstrated superior effectiveness in learning geometrical information, further validating its advantage in acoustic modeling.

\input{tab/table_ablation_1.tex}

\vspace{-1.5em}
\paragraph{Spectral feature refinement.}
To make realistic audio from obtained sound acoustics, we need to compensate for the time domain effect caused by time delay and reverberation and their corresponding spectral variations.
We explored a more compelling feature refinement network by replacing conventional architectures with a more expressive model.
We first compared several network designs, including a standard Convolutional network, U-Net~\cite{ronneberger2015u} with and without skip connections using a depth of 3, and ConvNeXt with a commonly used block of 3, incorporating modern convolutional improvements. Additionally, we compare with the introduced Multi-kernel ConvNeXt, which serves as the core architecture of the SRN in our AV-Surf model.
We provide refinement quality of each base network by visualizing spectrograms and corresponding waveforms in Fig.~\ref{fig:srn} with quantitative comparison in Table.~\ref{tab:ablation_2_a}.
Our proposed approach achieves the best performance shown in Table, demonstrating that capturing richer spectral representations leads to more realistic novel-view audio synthesis. To further analyze the impact of network depth, we conducted an ablation study on the number of Multi-Kernel ConvNeXt blocks, varying from 1 to 12.
The results in Table.~\ref{tab:ablation_2_b} indicate that increasing the number of blocks improves performance up to 9 blocks.
However, using excessive depth for our model yields a performance decrease.
Finally, the empirical results demonstrate that our SRN enhances binaural audio generation performance.

\input{tab/table_ablation_2.tex}

%% file: tab/rwavs_table.tex
\setlength{\tabcolsep}{7pt}
\begin{table*}[t]
\centering
\resizebox{\linewidth}{!}{
\begin{tabular}{l|cc|cc|cc|cc|cc}
\toprule
\multirow{3}{*}{\vspace{6pt} Methods} & \multicolumn{2}{c|}{Office} & \multicolumn{2}{c|}{House} & \multicolumn{2}{c|}{Apartment} & \multicolumn{2}{c|}{Outdoors} & \multicolumn{2}{c}{Overall} \\ 
\cmidrule(lr){2-3} \cmidrule(lr){4-5} \cmidrule(lr){6-7} \cmidrule(lr){8-9} \cmidrule(lr){10-11} 
& MAG$\downarrow$ & ENV$\downarrow$ & MAG$\downarrow$ & ENV$\downarrow$ & MAG$\downarrow$ & ENV$\downarrow$ & MAG$\downarrow$ & ENV$\downarrow$ & MAG$\downarrow$ & ENV$\downarrow$ \\
\midrule
Mono-Mono & 9.269 & 0.411 & 11.889 & 0.424 & 15.120 & 0.474 & 13.957 & 0.470 & 12.559 & 0.445\\
Mono-Energy & 1.536 & 0.142 & 4.307 & 0.180 & 3.911 & 0.192 & 1.634 & 0.127 & 2.847 & 0.160\\
Stereo-Energy & 1.511 & 0.139 & 4.301 & 0.180 & 3.895 & 0.191 & 1.612 & 0.124 & 2.830 & 0.159\\
\midrule 
INRAS~\cite{su2022inras} & 1.405 & 0.141 & 3.511 & 0.182 & 3.421 & 0.201 & 1.502 & 0.130 & 2.460 & 0.164\\
NAF~\cite{luo2022learning} & 1.244 & 0.137 & 3.259 & 0.178 & 3.345 & 0.193 & 1.284 & 0.121 & 2.283 & 0.157 \\
ViGAS~\cite{chen2023novel} & 1.049 & 0.132 & 2.502 & 0.161 & 2.600 & 0.187 & 1.169 & 0.121 & 1.830 & 0.150 \\ 
AV-NeRF~\cite{liang2023av} & 0.930 & 0.129 & 2.009 & 0.155 & 2.230 & 0.184 & 0.845 & \underline{0.111} & 1.504 & \underline{0.145} \\
AV-GS~\cite{bhosale2024av} & \underline{0.861} & \underline{0.124} & \underline{1.970} & \underline{0.152} & \underline{2.031} & \textbf{0.177} & \underline{0.791} & \textbf{0.107} & \underline{1.417} & \textbf{0.140} \\
\rowcolor{gray!20}\textbf{AV-Surf~(Ours)} & \textbf{0.760} & \textbf{0.123} & \textbf{1.934} & \textbf{0.151} & \textbf{1.999} & \underline{0.179} & \textbf{0.740} & \textbf{0.107} & \textbf{1.358} & \textbf{0.140} \\
\bottomrule
\end{tabular}
}
\vspace{-0.8em}
\caption{
Quantitative results on real-world RWAVS dataset.
Bold indicates the best, while the second-best is underlined.
}
\vspace{-1.8em}
\label{tab:rwavs}
\end{table*}

%% file: tab/soundspace_tab.tex
\setlength{\tabcolsep}{4pt} 
\begin{table}[t]
\centering
\resizebox{0.5\textwidth}{!}{
\begin{tabular}{l |c@{\hskip 3pt}c| c| c| c}
\toprule
\multirow{2}{*}{Methods} & \multicolumn{2}{c|}{Modality} 
 &  \multirow{2}{*}{T60 (\%) $\downarrow$}  & \multirow{2}{*}{C50 (dB) $\downarrow$} & \multirow{2}{*}{EDT (sec) $\downarrow$} \\

\multicolumn{1}{l|}{} & Audio \hspace{-3mm} & \multicolumn{1}{c|}{Visual} &  &  &   \\

\midrule
Opus-nearest  & \cmark & \xmark& 10.10 & 3.58 & 0.115 \\
    Opus-linear & \cmark & \xmark& 8.64  & 3.13 & 0.097 \\
    AAC-nearest & \cmark & \xmark& 9.35  & 1.67 & 0.059 \\
    AAC-linear & \cmark & \xmark& 7.88  & 1.68 & 0.057 \\
\hline
NAF  & \cmark & \xmark& 3.18  & 1.06 & 0.031 \\
    INRAS & \cmark & \xmark & 3.14  & 0.60 & 0.019 \\
\hline
AV-NeRF & \cmark & \cmark & 2.47 & 0.57 & 0.016 \\
AV-GS & \cmark & \cmark & \underline{2.23} & \underline{0.53} & \textbf{0.014} \\
\rowcolor{gray!20}\textbf{AV-Surf~(Ours)} & \cmark & \cmark & \textbf{2.04} & \textbf{0.52} &  \underline{0.015} \\
\bottomrule
\end{tabular}
}

\vspace{-0.8em}
\caption{
Quantitative results on synthetic Soundspaces dataset.
Bold indicates the best, while the second-best is underlined.}
\vspace{-1.6em}
\label{tab:soundspaces}
\end{table}

%% file: tab/table_ablation_1.tex
\begin{table}[t]
    \centering
    \begin{subtable}{0.23\textwidth}
    \centering
    \scalebox{0.9}{\scriptsize
    \begin{tabular}{l|c|c}
        \toprule
        \multirow{3}{*}{Methods} & \multicolumn{2}{c}{Overall} \\
        \cmidrule(lr){2-3}
        & MAG$\downarrow$ & ENV$\downarrow$ \\
        \midrule
        w/o surface normal  & 1.432 & 0.142 \\
        w/o pointcloud  & 1.386 & 0.141 \\
        \rowcolor{gray!20}\textbf{Ours} & \textbf{1.358} & \textbf{0.140} \\
        \bottomrule
        \end{tabular}}
        \caption{Feature type ablation}
        \label{tab:ablation_1_a}
    \end{subtable}
    \hspace{0.02em}
    \begin{subtable}{0.22\textwidth}
    \centering
    \scalebox{0.9}{\scriptsize
    \begin{tabular}{l|c|c}
        \toprule
        \multirow{3}{*}{Methods} & \multicolumn{2}{c}{Overall} \\
        \cmidrule(lr){2-3}
        & MAG$\downarrow$ & ENV$\downarrow$ \\
        \midrule
        MLP  & 1.372 & 0.141 \\
        CA  & 1.368 & 0.141 \\
        \rowcolor{gray!20}\textbf{Dual CA~(Ours)} & \textbf{1.358} & \textbf{0.140} \\
        \bottomrule
        \end{tabular}}
        \caption{Fusion method ablation}
        \label{tab:ablation_1_b}
    \end{subtable}
    \vspace{-0.8em}
    \caption{Ablation study for geometric priors effect.}
    \label{tab:ablation_1}
    \vspace{-2em}
\end{table}

%% file: tab/table_ablation_2.tex
\begin{table}[t]
    \centering
    \begin{subtable}{0.24\textwidth}
    \centering
    \scriptsize
    \begin{tabular}{l|c|c}
        \toprule
        \multirow{3}{*}{Methods} & \multicolumn{2}{c}{Overall} \\
        \cmidrule(lr){2-3}
        & MAG$\downarrow$ & ENV$\downarrow$ \\
        \midrule
        Convnet  & 1.410 & 0.142 \\
        U-net w/o skip  & 1.394 & 0.141 \\
        U-net w/ skip  & 1.380 & 0.141 \\
        ConvNeXt  & 1.373 & 0.141 \\
        \rowcolor{gray!20}\textbf{M.K-ConvNeXt} & \textbf{1.358} & \textbf{0.140} \\
        \bottomrule
        \end{tabular}
        \caption{Network type ablation}
        \label{tab:ablation_2_a}
    \end{subtable}
    \begin{subtable}{0.22\textwidth}
    \centering
    \scriptsize
    \begin{tabular}{l|c|c}
        \toprule
        \multirow{3}{*}{Methods} & \multicolumn{2}{c}{Overall} \\
        \cmidrule(lr){2-3}
        & MAG$\downarrow$ & ENV$\downarrow$ \\
        \midrule
        blk=1  & 1.378 & 0.141 \\
        blk=3  & 1.371 & 0.141 \\
        blk=6  & 1.368 & 0.141 \\
        \rowcolor{gray!20}\textbf{blk=9} & \textbf{1.358} & \textbf{0.140} \\
        blk=12  & 1.362 & 0.140 \\
        \bottomrule
        \end{tabular}
        \caption{Block number ablation}
        \label{tab:ablation_2_b}
    \end{subtable}
    \vspace{-0.8em}
    \caption{Ablation study for Sepctral feature refinement.}
    \label{tab:ablation_2}
    \vspace{-1.6em}
\end{table}

%% file: tex/conclusion.tex
\vspace{-0.4em}

\section{Conclusion}
\vspace{-0.8em}
In this work, we study how rich geometric visual cues enhance modeling the real-world acoustic. To learn realistic acoustic properties, we present AV-Surf, a comprehensive audio synthesis framework incorporating various geometry priors to encourage understanding of audio-visual modeling. Our proposed method enhances qualitative binauralization performance by utilizing surface normals derived from 2D Gaussian Splatting. Furthermore, our proposed SRN (Spectral Refinement Network) for synthesizing novel-view binaural audio captures spatial and spectral characteristics with fine-grained frequency-time details. Our method's qualitative and quantitative results demonstrate the quality of binauralization performance on real-world and synthetic datasets compared to existing approaches.
\vspace{-1.4em}

\paragraph{Limitations and Future Work.}
AV-Surf relies on the quality of geometric priors including surface normal and depth rendered from 2DGS. Reconstructing surface normal and depth from 3D representation learning are still developing, which can influence our visual features fusion process with geometric priors. Additionally, as the depth of our SRN block increases, the computational cost may also rise in proportion. We hope that further research on these challenges will be conducted, leading to greater improvements in final performance.
